\documentclass[9pt,twocolumn,twoside]{optica}
\setboolean{shortarticle}{false}
\setboolean{minireview}{false}

\newcommand{\rev}[1]{{\textcolor[rgb]{0.0,0.00,0.0}{#1}}}

\title{On-chip $\chi^{(2)}$ microring optical parametric oscillator}

\author[1]{Alexander W. Bruch}
\author[1]{Xianwen Liu}
\author[1]{Joshua B. Surya}
\author[2]{Chang-Ling Zou}
\author[1, *]{Hong X. Tang}

\affil[1]{Department of Electrical Engineering, Yale University, New Haven, CT 06511, USA}
\affil[2]{Department of Optics and Optics Engineering, University of Science and Technology of China, Hefei, Anhui 230026, China}
\affil[*]{Corresponding author: hong.tang@yale.edu}

\begin{abstract}
Optical parametric oscillators (OPOs) have been widely used for decades as tunable, narrow linewidth, and \rev{coherent light} sources for reaching long wavelengths and are attractive for applications such as quantum random number generation and Ising machines. \rev{To date, 
waveguide-based OPOs have suffered from relatively high thresholds on the order of hundreds of milliwatts.} With the advance in integrated photonic techniques demonstrated by high-efficiency second harmonic generation in aluminum nitride (AlN) photonic \rev{microring resonators}, highly compact and nanophotonic implementation of \rev{parametric oscillation} is feasible. \rev{Here, we employ phase-matched AlN microring resonators to demonstrate low-threshold parametric oscillation in the telecom infrared band with an on-chip efficiency up to 17\%  and milliwatt-level output power}. A broad phase-matching window is observed, enabling tunable generation of signal and idler pairs over a 180 nm bandwidth across the C band. This result establishes an important milestone in integrated nonlinear optics and paves the way towards chip-based quantum light sources and \rev{tunable, coherent radiation for spectroscopy and chemical sensing}.
\end{abstract}

\setboolean{displaycopyright}{true}

\begin{document}

\maketitle

\section{Introduction}
For decades, optical parametric oscillators (OPOs) have been a source for coherent radiation for reaching long wavelengths \cite{Tang1992, Godard2007, Breunig2011}. \rev{Traditional OPOs} rely on a $\chi^{(2)}$ material inserted in an \rev{optical} cavity that is resonant for the pump wavelength as well as the generated signal and/or idler wavelengths \cite{Giordmaine1965, Myers1997}. \rev{The generated signal and idler waves can then be tuned by controlling the phase-matching via a myriad of techniques, such as tuning the temperature or angle of the $\chi^{(2)}$ crystal \cite{Eckardt1991}, engineering a fan-out grating of the crystal poling \cite{Powers1998}, rotation of a diffraction grating \cite{Vainio2009}, electro-optic shaping of the parametric gain spectrum \cite{Gross2002}, or tuning of the pump wavelength \cite{Siltanen2010}}. \rev{Meanwhile, mirrorless OPOs without cavity enhancement were reported through careful engineering of counter-propagating waves in a periodically-poled crystal \cite{Canalias2007}}. \rev{Beyond long-wavelength coherent radiation, there has been interest in OPOs for generating optical squeezed states \cite{Wu1986}, correlated photon pair sources \cite{Morin2012, Fortsch2013, Vernon2015, Guo2016c, Lu2019}, and quantum random number generation \cite{Yamamoto2017}}.

Unlike second harmonic generation (SHG), OPO has a power threshold that is demanding on both the optical loss and modal phase-matching of the device \cite{Breunig2016b}. The first demonstration of microcavity-based \rev{OPO} was achieved in a bulk lithium niobate whispering gallery resonator (WGR), producing signal and idler pairs near 1100 nm from a 532 nm pump \cite{Furst2010a}. The very high optical quality (\emph{Q}) factors of \rev{$\mathrm{10^7 - 10^8}$} afforded by the WGR system, enabled an \rev{OPO} threshold of 6 $\mu \mathrm{W}$ \cite{Furst2010a}. Radially or linearly poling these devices further enabled tunable quasi-phase-matching beyond 2 $\mu \mathrm{m}$ wavelengths with in-resonator OPO conversion efficiencies near 50\% \cite{Beckmann2011, Werner2012, Beckmann2012, Breunig2013, Herr2018}. Recent work has employed \rev{OPOs} for high resolution spectroscopy in the mid-infrared regime \cite{Meisenheimer2015, Werner2015} and has extended the wavelength of the idler waves beyond 8\,$\mu \mathrm{m}$ using novel materials such as $\mathrm{AgGaSe_2}$ and $\mathrm{CdSiP_2}$ \cite{Jia2018, Meisenheimer2017}.

\rev{Despite progresses in bulk WGRs, low-threshold OPOs have proven to be quite challenging in planar photonic platforms. A milliwatt-level OPO threshold was observed in a planar periodically-poled Ti:$\mathrm{LiNbO_3}$ waveguide, requiring a relatively long crystal length of 80 mm \cite{Schreiber2001}. Monolithic semiconductor OPOs were later realized in orientation patterened GaAs \cite{Oron2012} as well as GaAs/AlGaAs \cite{Savanier2013} ridge waveguides by employing dielectric coatings on the chip facets. However, these devices suffered from relatively large OPO thresholds of 5.7 W and 210 mW, respectively.} 
Recently, epitaxial aluminum nitride (AlN) has emerged as a compelling photonic platform with the achievement of high-Q microring resonators from the ultraviolet to near-infrared wavelengths \cite{Liu2018a,Liu2017} and the demonstration of efficient $\chi^{(2)}$ and $\chi^{(3)}$  nonlinear processes \cite{Liu2019, Liu2017d, Liu2018}. A recent demonstration of a record-high SHG conversion efficiency of 17,000 \%/W \cite{Bruch2018} further poses AlN as a serious contender to traditional $\chi^{(2)}$ materials such as lithium niobate (LN). While the intrinsic Pockels coefficient of AlN ($\sim$6 pm/V \cite{Bruch2018}) is less than that of lithium niobate ($\sim$30 pm/V \cite{Zhang2017}), it is free of photo-refractive effects and is less susceptible to two-photon absorption losses, making it a promising material platform for \rev{low-threshold, chip-integrated optical parametric oscillation}.

\rev{In this paper, we present the first demonstration of optical parametric oscillation in a waveguide-integrated AlN microring resonator}. By optimizing the modal phase-matching and dual-resonance condition between the near-visible (780 nm) and the telecom infrared (IR) band (1560 nm), we achieve a low OPO threshold of 12 mW as well as 21\% SHG and 17\% OPO power conversion efficiencies. We show that the phase-matching condition can be well controlled via an external heater, which allows the signal and idler pair to be tuned across a bandwidth of 180 nm (23 THz) including the final transition from non-degenerate to degenerate OPO. Our approach can be extended to achieve chip-based, \rev{narrow-linewidth light sources} at other wavelengths important for a variety of potential applications including molecular spectroscopy and chemical sensing, as well as in other $\chi^{(2)}$ materials such as LiNbO$_3$ on insulator \cite{Zhang2017, Desiatov2019}. 

\section{Modeling of the OPO threshold and parametric oscillation behavior}

\begin{figure}[htb] 
    \centering
    \includegraphics[width=\linewidth]{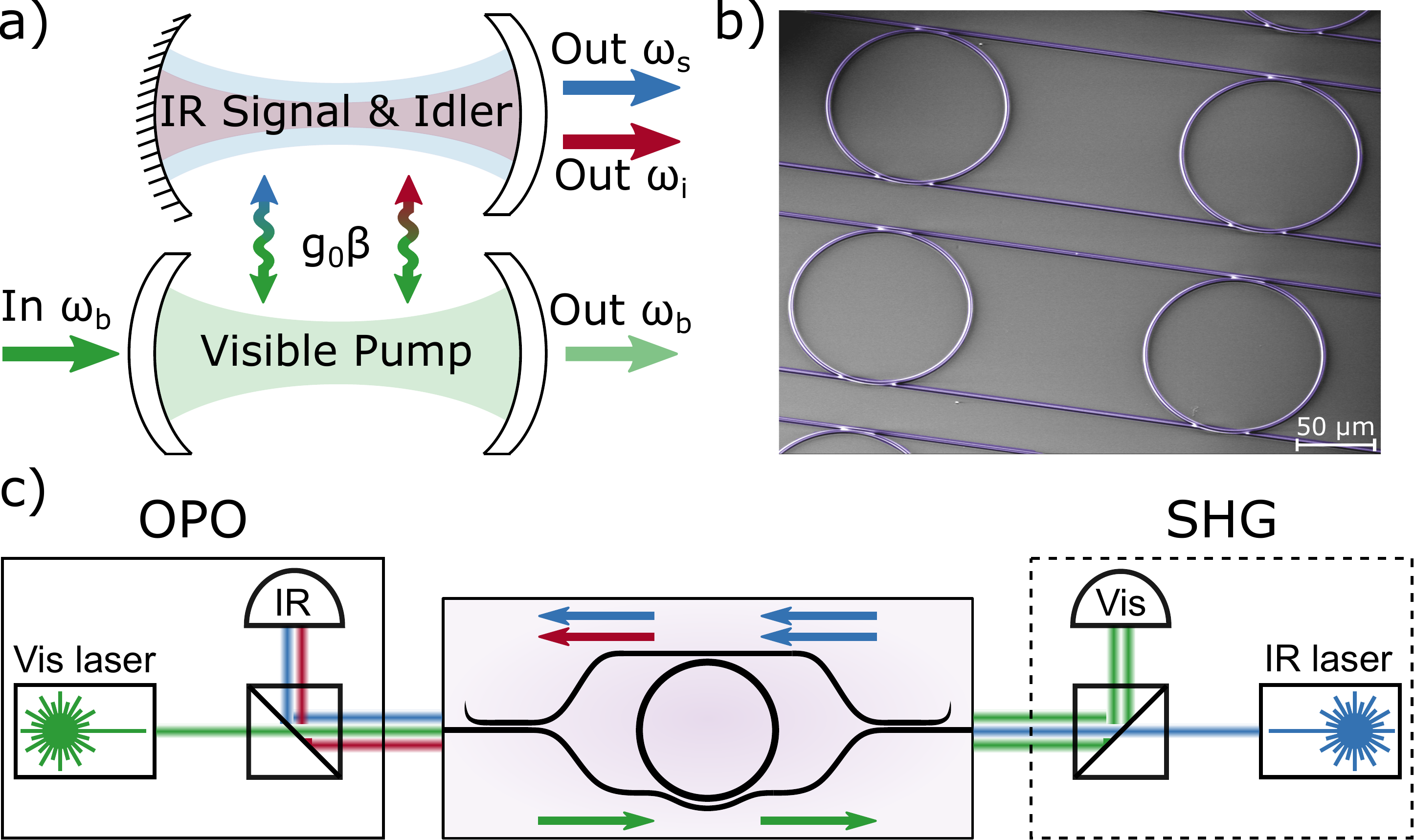}
    \caption{(a) Schematic representation of the \rev{parametric oscillation} model using two Fabry-Per\'ot cavities. The visible mode near 780 nm ($\omega_b$, green) produces nonlinear gain in the infrared signal and idler modes near 1560 nm  ($\omega_s$ and $\omega_i$, blue and red, respectively) via the $\chi^{(2)}$ effect with strength $g_0 \beta$. (b) Colorized scanning electron microscope (SEM) image of the fabricated AlN chip with cascaded microring resonators before SiO$_2$ encapsulation. (c) Schematic of the OPO (left) and SHG (right) measurement schemes. On-chip, the top bus waveguide addresses the infrared modes (red and blue) while the bottom bus waveguide addresses the near-visible modes (green). Note that in both cases the SHG and OPO waves are collected from the input facet of the chip. The infrared and near-visible waves are separated by an off-chip WDM before detection (shown here as a dichroic beamsplitter).} 
    \label{fig:cavities}
\end{figure}

Figure \ref{fig:cavities} (a) schematically illustrates the \rev{parametric oscillation process}, where the system under study can be idealized as two \rev{coupled cavities}. A visible pump laser at an angular frequency $\omega_b$ pumps the device, producing a signal and idler pair at frequencies $\omega_s$ and $\omega_i$ (blue and red, respectively) which satisfies the energy matching condition $\omega_s + \omega_i = \omega_b$. For the degenerate OPO process ($\omega_s = \omega_i = \omega_b/2$), a single frequency \rev{oscillation} is realized at half the frequency (twice the wavelength) of the pump. In the non-degenerate case, $\omega_s \neq \omega_i$ and \rev{parametric oscillation} is realized at two distinct resonances centered about the pump.  

In contrast to the derivation from Refs. \cite{Sturman2011, Breunig2016b}, we derive the OPO threshold via a bosonic coupled-mode model, which is also applicable for other $\chi^{(2)}$ processes such as spontaneous parametric down-conversion (SPDC) and on-chip strong coupling \cite{Guo2016b, Guo2016c, Guo2018a}. \rev{A similar Hamiltonian approach for modeling OPO in a WGR can be found in Ref. \cite{Ilchenko2003}}. To gain insights in the OPO process and a comparison to SHG, here we focus the derivation on degenerate OPO. Details on non-degenerate OPO modeling is provided in Supplementary Information. 

As in Ref.~\cite{Guo2016b}, the total Hamiltonian of degenerate OPO and SHG reads 
\begin{equation}
    \mathcal{H}/\hbar =  \omega_a a^{\dagger}a +  \omega_b b^{\dagger}b  +  g_0b(a^\dagger)^2 +  g_0 a^2 b^\dagger,
    \label{Htot}
\end{equation}
where $\omega_a$ represents the mode angular frequency for the infrared modes $a$, $\omega_b$ is the mode angular frequency for near-visible mode $b$ and $g_0$ is the nonlinear coupling strength between modes $a$ and $b$. We then apply an external \rev{pump} laser \rev{near mode $b$} at a frequency $\omega_p$ with the strength
\begin{equation}
    \beta = \frac{\sqrt{2\kappa_{b,1}}}{-i(\omega_b - \omega_p)-\kappa_b}\sqrt{\frac{P_p}{\hbar \omega_p}}
    \label{beta}
\end{equation}
where $P_p$ is the pump power and $\kappa_{a(b)} = \kappa_{a(b),0} + \kappa_{a(b),1}$ is the total amplitude decay rate of mode $a(b)$, with subscripts {0,1} denoting the intrinsic and external coupling losses, respectively. Applying a mean field approximation in the rotating frame of $\omega_{a}$, the effective Hamiltonian at infrared probe mode $a$ becomes
\begin{equation}
    \mathcal{H}_{eff}/\hbar = \delta_a a^\dagger a + g_0 \beta ((a^{\dagger})^2 + a^2)
    \label{Heff}
\end{equation}
where $\delta_a = \omega_a - \omega_p/2$ represents the angular frequency detuning of the signal (idler) from the down-converted pump. 

We note that the term $g_0 \beta$ in Eq.~(\ref{Heff}) denotes the nonlinear gain of photons at mode $a$ when driven by the pump mode $b$. The mode $a$ will begin to \rev{oscillate} when the nonlinear gain $g_0\beta$ is greater than its total cavity loss given by
\begin{equation}
    \kappa_a^2 \leq g_0^2\beta^2 = g_0^2\frac{2\kappa_{b,1}}{(\omega_b-\omega_p)^2 + \kappa_b^2}\frac{P_{th}}{\hbar \omega_p}.
    \label{opo_thresh_general}
\end{equation}
A full derivation of the \rev{parametric oscillation} condition can be found in the Supplementary Information. Using $\kappa_{a(b)} = \frac{\omega_{a(b)}}{2Q_{a(b)}}$ and assuming critical coupling ($\kappa_{a(b),1} = \kappa_{a(b),0} = \kappa_{a(b)}/2$) \rev{and the external pump on resonance with mode $b$ ($\omega_p = \omega_b$)}, the OPO threshold power can be derived as

\begin{equation}
    P_{th} = \frac{\hbar \omega_b}{g_0^2} \kappa_{a,0}^2 \kappa_{b,0} = \frac{\hbar \omega_b^4}{32 g_0^2} \frac{1}{Q_{a,0}^2 Q_{b,0}} 
    \label{opo_thresh_power}
\end{equation}
Here, $Q_{a(b),0}$ is intrinsic intrinsic  quality factor of mode $a(b)$. Compared with the SHG efficiency $\eta_{SHG}$ below \cite{Guo2016, Bruch2018}

\begin{equation}
    \eta_{SHG} = 
    \frac{P_{b}}{P_a^2} = \frac{g^2}{4 \kappa_{a,0}^2 \kappa_{b,0}} \frac{1}{\hbar \omega_a} = \frac{g_0^2 Q_{a,0}^2 Q_{b,0}}{\hbar \omega_a^4}
    \label{eta_shg}
\end{equation}
we find that $P_{th} =\frac{2}{\eta_{SHG}}$ after assuming $\omega_b = 2\omega_a$, suggesting a lower OPO threshold for a device with a higher SHG efficiency.

Above the OPO threshold, we solve Eq.~(\ref{Heff}) in the steady-state ($\dot{a} = \dot{b} = 0$) to find
\begin{equation}
    |a|^2 = \frac{\sqrt{2 \kappa_{b,1}}}{g_0} \sqrt{\frac{P_b}{\hbar \omega_b}} - \frac{\kappa_a \kappa_b}{2 g_0^2}.
    \label{a_amp}
\end{equation}
Defining the single-photon cooperativity
\begin{equation}
    C_0 = \frac{g_0^2}{\kappa_a \kappa_b} = \frac{1}{\frac{P_{th}}{\hbar \omega_b} \frac{8 \kappa_{b,1}}{\kappa_b} \frac{1}{\kappa_a}}
\end{equation}
we can simplify Eq.~(\ref{a_amp}) to $|a|^2 = \frac{1}{C_0} (\sqrt{P_b/P_{th}} - 1)$. The total OPO output power then reads

\begin{eqnarray}
 P_{s+i} & = & 2 \kappa_{a,1} \hbar \omega_a |a|^2\\
 & = & 8 \frac{\kappa_{a,1}}{\kappa_a} \frac{\kappa_{b,1}}{\kappa_b} P_{th}(\sqrt{P_b/P_{th}} - 1).
\label{P_opo}
\end{eqnarray}
and the corresponding OPO efficiency is
\begin{equation}
    \eta_{s+i} = \frac{P_{s+i}}{P_P} = 8 \frac{\kappa_{a,1}}{\kappa_a} \frac{\kappa_{b,1}}{\kappa_b} \frac{\sqrt{P_b P_{th}}  - P_{th}}{P_b}.
    \label{eta_opo}
\end{equation}
A full derivation of these equations and their equivalence to those derived in Ref.~\cite{Breunig2016b} are given in the Supplementary Information.

\section{Device Fabrication and Measurement}

Figure~\ref{fig:cavities}(b) highlights the fabricated single-crystalline AlN device specifically designed for SHG and OPO. The principal components include a microring resonator with an optimized width of 1.20 $\mu$m for phase-matching between 780 nm and 1560 nm as well as two bus waveguides for separately addressing each mode. The optimal visible and infrared coupling gaps are 0.4 and 0.7 $\mu$m, respectively. Meanwhile, we adopt a relatively large microring radius of 60 $\mu \mathrm{m}$ to minimize the radiation loss of the infrared signal and idler. The outgoing infrared and near-visible light is then separated via an on-chip wavelength division multiplexer (WDM) (not shown) \cite{Guo2016c}. A weakly tapered pulley-like coupler is used to enhance the coupling strength of the visible mode with minimal loss on the infrared mode \cite{Guo2016, Guo2016a, Guo2016c, Bruch2018}.

The device fabrication begins with an AlN thin film (thickness of 1.0 $\mu$ m) epitaxially grown on c-plane (0001) sapphire substrate by metal-organic chemical vapor deposition. The patterns are then defined by electron beam lithography and transferred to the AlN film using a $\mathrm{Cl_2/BCl_3/Ar}$-based inductively-coupled plasma etch. After encapsulation within silicon dioxide ($\mathrm{SiO_2}$) by plasma-enhanced chemical vapor deposition, the AlN chip is cleaved for characterization. The full fabrication process is presented in Ref. \cite{Bruch2018}.

Figure \ref{fig:cavities} (c) schematically shows our experimental setup for characterizing the SHG and OPO processes. For SHG (OPO), the device is pumped by an IR (visible) laser and the visible (IR) power is collected on the corresponding photodetector. In each case, the SHG or OPO light is collected from the input facet of the AlN chip. An off-chip wavelength division multiplexer (WDM) separates the visible and infrared light before collecting into a photodetector. The two measurements are conducted separately, indicated by the solid and dashes boxes in Fig. \ref{fig:cavities}(c).

\section{Results and Discussion}
\subsection{Second Harmonic Generation}
\begin{figure}[ht] 
    \centering
    \includegraphics[width=\linewidth]{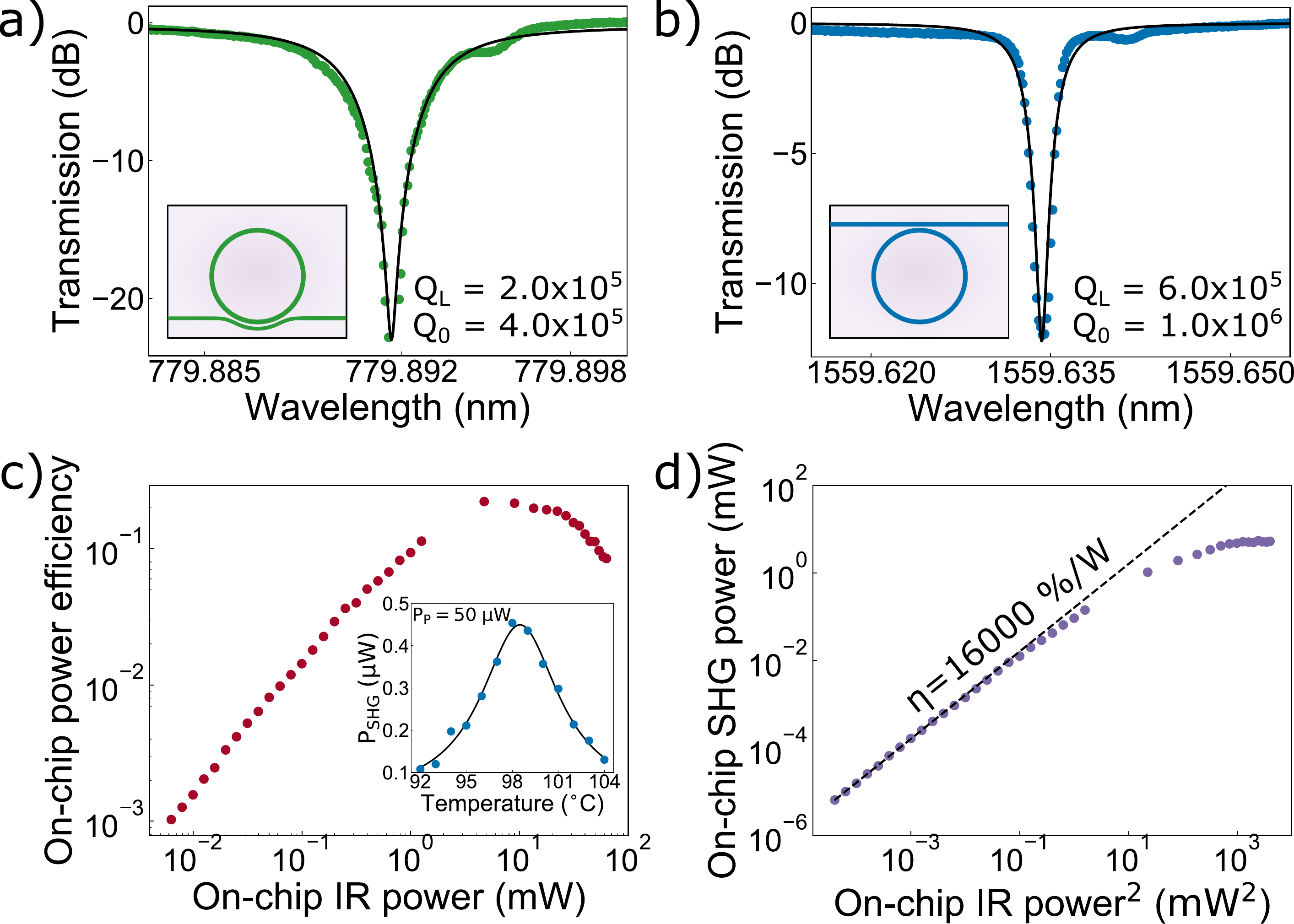}
    \caption{Measured resonance spectra for the visible (a) and infrared (b) modes. The loaded and intrinsic \emph{Q} factors ($Q_L$ and $Q_0$, respectively) are extracted after applying a Lorentzian fit at undercoupled conditions. The insets highlight the bus waveguide addressing each resonance. (c) On-chip SHG power conversion efficiency ($\mathrm{P_b/P_a}$) vs on-chip IR pump power. Inset: temperature dependence of the maximum on-chip SHG power (blue dots) with on-chip pump power of 50 $\mu$W. A Lorentzian fit (black line) is applied to determine the optimum temperature. (d) $\mathrm{P_b}$ versus $\mathrm{P_a^2}$, where a linear fit in the low power regime is used to extract the SHG conversion efficiency ($\mathrm{P_b/P_a^2}$). The break of the data in (c) and (d) occurs when an erbium-doped fiber amplifier is introduced to provide high pump power.}
    \label{fig:shgdata}
\end{figure}

We first characterized the visible and infrared optical \emph{Q}-factors of the AlN resonator by scanning the optical resonances with a New Focus TLB-6712 (visible) and Santec TSL 710 (infrared) lasers, respectively. As shown in Fig. \ref{fig:shgdata}(a), the visible and infrared resonances were observed to have intrinsic optical \emph{Q} factors of $\sim$400 k and $\sim$1.0 million, respectively. The visible resonance is nearly critically coupled with an extinction greater than 20 dB whereas the infrared resonance is slightly under-coupled with an extinction of 12 dB to reduce the OPO threshold \cite{Breunig2016b}.\rev{The coupling condition of the resonators is examined by varying the coupling gap between the bus waveguide and ring resonator}.

It is known that high-efficiency \rev{OPO} in a microcavity requires phase-matching as well as a dually-resonant condition for the involved infrared and visible modes, akin to highly efficient SHG. For this purpose, we first optimized the dual-resonance condition by tuning the temperature of the chip via an external heater. As shown in the inset to Fig. \ref{fig:shgdata}(c), a maximum on-chip SHG power of 0.45 $\mu$W (infrared pump of  50 $\mu$W) was observed at an optimal microring width of 1.2 $\mu$m and a heater temperature of 98 $\mathrm{^\circ C}$. 

The pump power at this optimized temperature was subsequently varied to determine the maximum power efficiency ($\mathrm{P_b/P_a}$). Figure \ref{fig:shgdata}(c) shows that the power efficiency increased rapidly in the low power regime and saturates to a maximum value of 21\% at higher on-chip pump powers above 1 mW and finally decreases due to pump depletion \cite{Furst2010}. The SHG conversion efficiency ($\mathrm{P_b/P_a^2}$) is shown in \rev{Fig. \ref{fig:shgdata}(d)}, where a linear fit is applied in the low-power region, revealing a mean SHG efficiency of 16,000 \%/W. 

The observed power efficiency ($\mathrm{P_b/P_a}$) is double our previous result in Ref.~\cite{Bruch2018}, and is close to the theoretical maximum for a critically coupled SHG resonance and a slightly under-coupled pump resonance \cite{Sturman2011}, while the SHG efficiency ($\mathrm{P_b/P_a^2}$) remained comparable. Most importantly, the results of the SHG experiment allow us to calculate the key parameter $g_0$ for estimating the OPO threshold. Based on the measured  $\kappa_a$ and $\kappa_b$ in Figs. \ref{fig:shgdata}(a) and \ref{fig:shgdata}(b), we extract an experimental $g_0/2\pi \approx$ 80 kHz from Eq.~(\ref{eta_shg}) and estimate an on-chip OPO threshold of $\sim$11 mW in this system according to Eq.~(\ref{opo_thresh_power}).

\subsection{Optical Parametric Oscillation and Tunability}

The AlN device with an optimized SHG efficiency in Fig. \ref{fig:shgdata}, was subsequently probed by varying the pump power from a Ti-sapphire laser (M2 SolsTis, 700$-$1000 nm) to investigate the OPO threshold. The details of the experimental setup are shown in the Supplementary Information.

\begin{figure}[htbp]
    \centering
    \includegraphics[width=\linewidth]{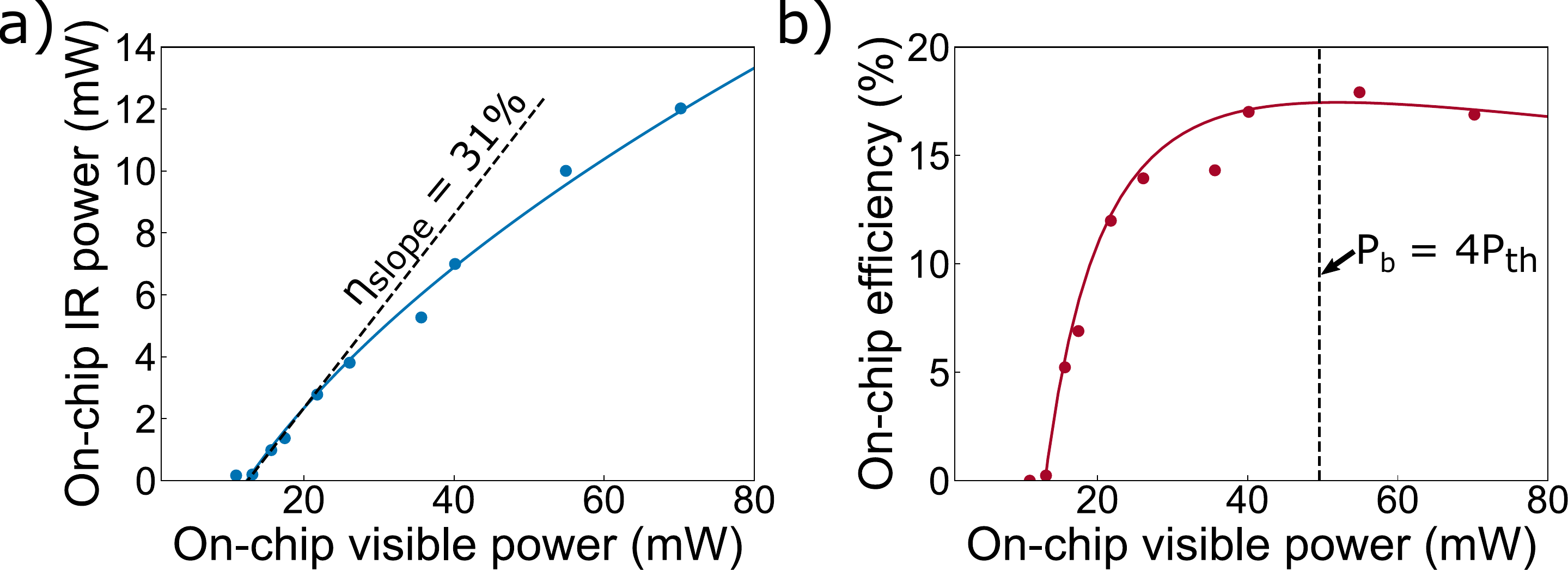}
    \caption{(a) Total on-chip infrared power versus on-chip visible pump power.  The data is fitted to the theory to determine the OPO threshold (blue line). Near the threshold, a linear fit (dashed line) is also applied to give a slope efficiency of 31\%. (b) Measured on-chip OPO conversion efficiency versus on-chip visible pump power, where a theoretical fit (red line) is applied to the data. The dashed line indicates the point at which the OPO efficiency has a theoretical maximum at $\mathrm{P_b=4P_{th}}$}
    \label{fig:shg_opo}
\end{figure}

Figure~\ref{fig:shg_opo}(a) plots the collected OPO power while varying the visible input power at the previously optimized phase-matching temperature of 98 $\mathrm{^\circ C}$. Infrared power was detected in an off-chip integrating sphere detector when the on-chip visible pump power was above $\sim$15 mW. By elevating the on-chip visible pump power to 70.2 mW, we record an off-chip OPO power of $\sim$4.5 mW, corresponding to an on-chip infrared power of 12.0 mW. The experimental results show good agreement with the theoretical fit  using Eq. \ref{P_opo} (solid blue line), where an OPO threshold of 12.3 mW is derived,  consistent with the estimated value from the SHG experiment (11 mW). A linear fit of the on-chip OPO power near the threshold also yields a high slope efficiency of 31\%. 

By calibrating the collected infrared light to the waveguide coupling efficiency (14\% and 41\% per facet at 780 and 1560 nm, respectively), we plot the on-chip conversion efficiency in Fig.~\ref{fig:shg_opo}(b), and fit Eq.~(\ref{eta_opo}) to the results (solid red lines). A maximum on-chip conversion efficiency of 17\% was achieved, which is quite close to half the slope efficiency in Fig.~\ref{fig:shg_opo}(a). We note that the maximum OPO efficiency in the experiment was realized at $\mathrm{P_b = 4.5 P_{th}}$, which is close to the theoretical value of $\mathrm{P_b = 4.0 P_{th}}$, indicated by the dashed line in Fig.~\ref{fig:shg_opo}(b). The agreement between the OPO threshold calculated from the SHG experiment (11 mW), the experimental OPO threshold (15 mW), and the value derived from the theoretical fit (12 mW) suggests that our analysis outlined above sufficiently describes the $\chi^{(2)}$ frequency conversion processes of our system.

\begin{figure}[ht]
    \centering
    \includegraphics[width=\linewidth]{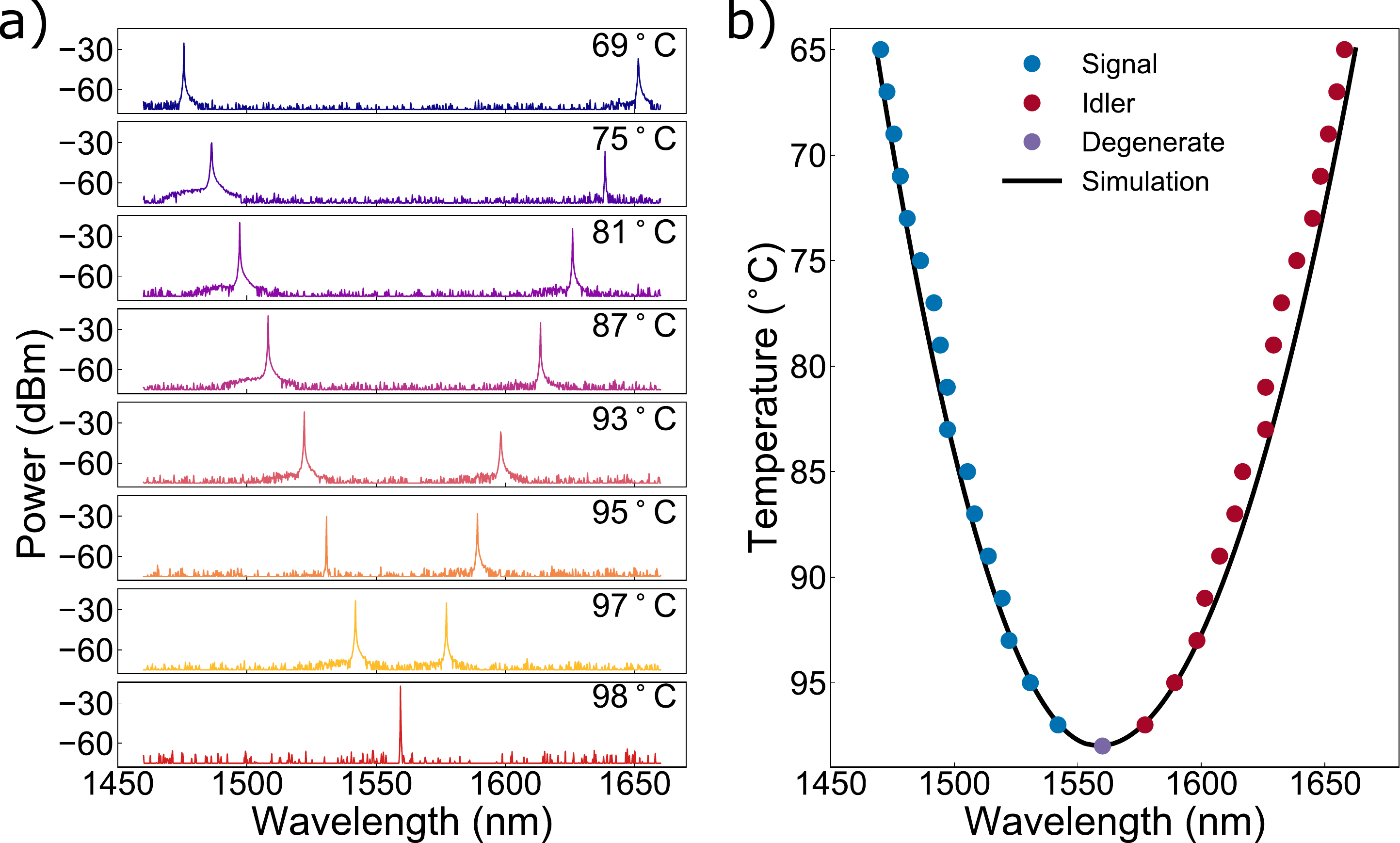}
    \caption{(a) \rev{OPO} spectra collected from the transmission port of the microring resonator at different temperatures. Degenerate \rev{OPO} occurs at 98 $\mathrm{^\circ C}$ (bottom), below which non-degenerate \rev{OPO} is observed (top). (b) Recorded \rev{OPO} wavelength versus the temperature. Degenerate \rev{parametric oscillation} (purple) occurs at the optimum temperature for SHG of 98 $\mathrm{^\circ C}$. The separation of the signal (blue) and idler (red) increases as the temperature is tuned away from this point. A numerical simulation of Eq. (\ref{thermal_index}) is displayed as a black line, consistent with the experimental result.}
    \label{fig:tuning}
\end{figure}

The full tuning bandwidth of the OPO was further investigated by controlling the phase-matching via temperature tuning \cite{Beckmann2011, Meisenheimer2017}. We begun at the optimal phase-matching temperature of 98 $\mathrm{^\circ C}$ and proceeded to decrease the temperature while tuning the pump laser into the resonance. Moderate on-chip pump powers were applied ($\mathrm{P_b \sim }$ 30 mW) to ensure \rev{parametric oscillation} throughout the entire temperature tuning process. As shown in Fig. \ref{fig:tuning}(a), \rev{we observed degenerate parametric oscillation at 98 $\mathrm{^\circ C}$ as optimized in the SHG experiment}. Here, $\lambda_s = \lambda_i = 2\lambda_b$, where $\lambda_s$, $\lambda_i$, and $\lambda_b$ are the wavelengths of signal, idler, and pump, respectively. As the temperature was decreased from this optimum point, we observed the onset of non-degenerate \rev{oscillation} ($\lambda_s \neq \lambda_i$). The separation between the signal and idler was observed to increase with the decreasing temperature due to the thermal dependence of the effective refractive index, which alters the wavelengths that simultaneously satisfy energy and momentum conservation, given by

\begin{equation}
\begin{gathered}
    \frac{1}{\lambda_b} = \frac{1}{\lambda_s} + \frac{1}{\lambda_i} \\
    \frac{n_b(T)}{\lambda_b} = \frac{n_s(T)}{\lambda_s} + \frac{n_i(T)}{\lambda_i}.
\end{gathered}
\label{thermal_index}
\end{equation}

By varying the temperature, a range of $\lambda_s$ and $\lambda_i$ are available to satisfy the phase-matching condition. The full tuning range of the \rev{OPO} process is summarized in Fig. \ref{fig:tuning}(b), where a maximum span of 180 nm (25 THz) was achieved. The observed temperature tuning is also in good agreement with a numerical simulation of Eq.~(\ref{thermal_index}) (black line in Fig. \ref{fig:tuning}(b)) using experimentally extracted temperature-dependent effective indices $\frac{\partial n}{\partial T}$. We note that while the phase-matching condition can accommodate many values of $\lambda_s$ and $\lambda_i$, \rev{OPO} wavelengths must be commensurate with  cavity resonances and thus the tuning can only be varied by the cavity FSR ($\sim$2.5 nm). Nevertheless, the 180 nm tuning range of the signal and idler pair can be easily controlled with a bandwidth comparable to a commercial tunable laser. The linewidth of the degenerate and non-degenerate \rev{OPOs} were also estimated via delayed self-heterodyne beat note measurement as detailed in the Supplementary Information.

\begin{figure}[htb]
    \centering
    \includegraphics[width=0.9\linewidth]{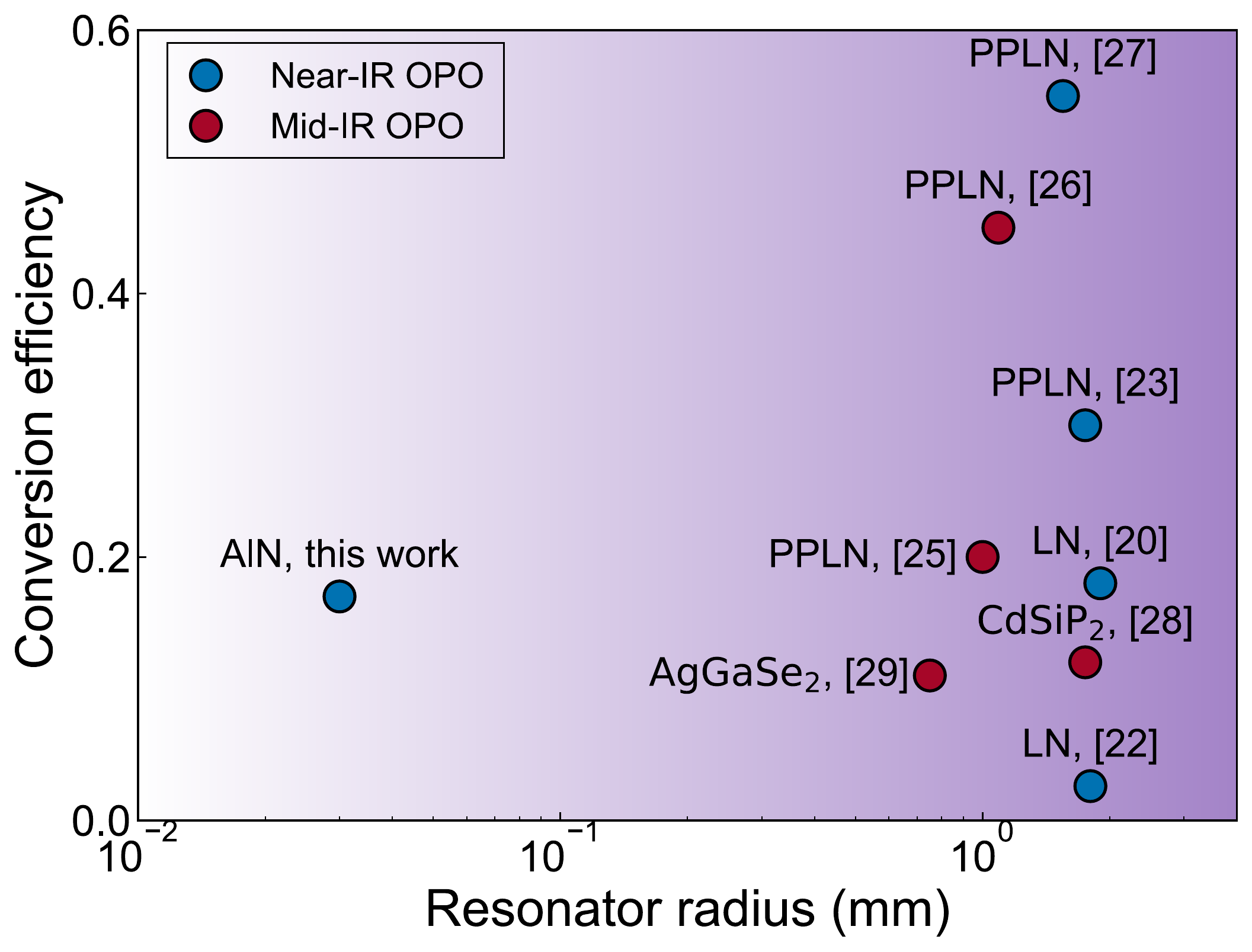}
    \caption{Pump-to-OPO conversion efficiencies reported for various \rev{microcavity-based} OPO devices. Blue and red circles represent OPO wavelengths in the near-infrared (<2 $\mu$m) and mid-infrared (>2 $\mu$m) regimes, respectively. LN: lithium niobate; PPLN: periodically-poled lithium niobate.}
    \label{fig:opo_comparison}
\end{figure}

We note the presented device exhibited a higher OPO threshold than that of previous LN bulk WGRs \cite{Beckmann2011, Werner2012, Beckmann2012, Breunig2013}, which we attribute to the limited optical \emph{Q} of our micro-scale device. Our AlN microring resonators are able to observe \rev{parametric oscillation} with relatively small optical \emph{Q} factors compared to bulk resonators due to their significantly reduced mode volume. The observed efficiency in our waveguide-integrated microring resonators is comparable to many of the previously demonstrated WGR devices as highlighted in Fig. \ref{fig:opo_comparison}. Meanwhile, our AlN microring resonator yielded a comparable threshold and efficiency compared to novel mid-IR OPO materials such as $\mathrm{AgGaSe_2}$ and $\mathrm{CdSiP_2}$ \cite{Meisenheimer2017, Jia2018}. \rev{Moreover, our AlN microring structure is able to reduce the OPO threshold one order of magnitude relative to previous GaAs/AlGaAs integrated waveguides without additional fabrication of dielectric mirrors on the chip facets \cite{Savanier2013}}.
The superior band-gap of AlN from the ultraviolet to mid-Infrared as well as the robustness of our chip-integrated system opens new routes for chip-scale, efficient OPOs from the near- to mid-infrared. 

\section{Conclusions}

To the best of our knowledge, this work demonstrates the first experimental observation of $\chi^{(2)}$ OPO in a waveguide-integrated microring resonator. Our microring-based OPO exhibits a high efficiency of 17\% and milliwatt-level off-chip output power. The observance of a broad tuning range of the generated signal and idler photons enables quasi-continuous tuning over a 180 nm (25 THz) band in the telecomm IR regime. While the current device has a higher threshold compared to previous bulky periodically-poled lithium niobate WGR structures, our nanophotonic platform features improved compactness and scalability as well as ease of design and fabrication.

Since AlN is also a viable platform for low-loss mid-infrared photonics \cite{Hickstein2017, Dong2019}, our approach is promising to facilitate nanophotonic chip-based \rev{OPOs} for mid-infrared applications, where very few narrow-linewidth, tunable, solid state sources are available. 
By moving the pump wavelength to the telecomm band, we anticipate the possibility to develop \rev{OPOs} above 3 $\mu$m, providing a tunable alternative to quantum cascade laser devices in this regime. Our approach is also applicable to other $\chi^{(2)}$ photonic material platforms, particularly thin film lithium niobate \cite{Zhang2017, Desiatov2019} and gallium arsenide \cite{Chang2018, Chang2019}, with large $\chi^{(2)}$ nonlinearities, which could further reduce the OPO threshold to sub-milliwatt levels in nanophotonic devices.

\section*{Funding Information}
This work is supported by DARPA SCOUT (W31P4Q-15-1-0006). H.X. Tang acknowledges partial support from DARPA's ACES programs as part of the Draper-NIST collaboration (HR0011-16-C-0118), an AFOSR MURI grant (FA9550-15-1-0029), a LPS/ARO grant (W911NF-14-1-0563), a NSF EFRI grant (EFMA-1640959) and Packard Foundation.   

\section*{Acknowledgments}


The facilities used for device fabrication were supported by the Yale SEAS Cleanroom and the Yale Instutite for Nanoscience and Quantum Engineering. The authors thank Dr. Yong Sun, Dr. Michael Rooks, Sean Rinehart, and Kelly Woods.


\bigskip \noindent See \href{link}{Supplement 1} for supporting content.



 

\bibliography{revised_references}



\end{document}